\documentclass[10.75pt]{article}
\usepackage{geometry}
\usepackage{makeidx}
\usepackage[T1]{fontenc}
\usepackage[dvipsnames,svgnames,table]{xcolor}
\usepackage{epstopdf}
\usepackage{epsfig}
\usepackage{textcomp}
\usepackage{hyperref}
\usepackage{amsmath}
\usepackage{amssymb}
\usepackage{multirow}
\usepackage{multicol}
\usepackage{booktabs}
\usepackage{lastpage}
\usepackage{hyperref} 
\usepackage{graphicx}
\usepackage{enumerate}
\usepackage{threeparttable}
\usepackage{float}
\usepackage{array}
\usepackage{cite}
\usepackage{color,soul}
\makeatletter
\renewcommand\section{\@startsection{section}{1}{\z@}%
  {-2.5ex \@plus -1ex \@minus -.2ex}%
  {2.3ex \@plus.2ex}%
  {\normalfont\large\bfseries}}
\renewcommand\subsection{\@startsection{subsection}{1}{\z@}%
  {-2.5ex \@plus -1ex \@minus -.2ex}%
  {2.3ex \@plus.2ex}%
  {\small\bfseries}}
\begin{document}
\title{\vskip -2.5 cm\textbf{On muon energy group structure based on deflection angle for application in muon scattering tomography: A Monte Carlo study through GEANT4 simulations}}
\medskip
\author{\small A. Ilker Topuz$^{1,2}$, Madis Kiisk$^{1,3}$, Andrea Giammanco$^{2}$, and Mart Magi$^{3}$}
\medskip
\date{\small$^1$Institute of Physics, University of Tartu, W. Ostwaldi 1, 50411, Tartu, Estonia\\
$^2$Centre for Cosmology, Particle Physics and Phenomenology, Universit\'e catholique de Louvain, Chemin du Cyclotron 2, B-1348 Louvain-la-Neuve, Belgium\\
$^3$GScan OU, Maealuse 2/1, 12618 Tallinn, Estonia}
\maketitle
\begin{abstract}
The average deflection angle of the tracked muons in the muon scattering tomography exponentially declines in function of the initial kinetic energy, the angular dependence of which provides an opportunity to set out a binary relation between the initial kinetic energy and the average deflection angle, thereby leading to a coarse energy prediction founded on the mean deflection angle in the case of experimental incapabilities or limitations. Nevertheless, in addition to the disadvantageous exponential trend, the standard deviations observed in the deflection angles restrict the number of energy groups by yielding a significant number of coincided angular outcomes even at the fairly distinct energy values. In this study, we address the problem of the muon energy classification for a tomographic system consisting of 0.4-cm plastic scintillators manufactured from polyvinyl toluene and we explore a four-group structure besides a ternary partitioning between 0.25 and 8 GeV. In the first instance, we determine the deflection angles by tracking the hit locations in the detector layers on the sub-divided uniform energy intervals through the GEANT4 simulations. In the latter step, we express two misclassification probabilities where the first approach assumes a symmetrical linear propagation bounded by one standard deviation in one dimension, whereas the second procedure employs a positively defined modified Gaussian distribution that governs the overlapping area in two dimensions. In the final stage, we compare qualitatively and quantitatively the adjacent energy groups by using the computed misclassification probabilities. In the absence of any further data manipulation, we explicitly show that the misclassification probabilities increase when the number of energy groups augments. Furthermore, we also conclude that it is feasible to benefit from the mean deflection angle to roughly estimate the muon energies up to four energy groups by taking the misclassification probabilities into consideration, while the classification viability significantly diminishes when the partition number exceeds four on the basis of standard deviation.
\end{abstract}
\section{Introduction}
Muon tomography is a relatively novel imaging technique~\cite{bonechi2020atmospheric} that makes use of the free natural flux of muons originating from the interaction of cosmic rays in the atmosphere. One of its classes of applications is the material discrimination (e.g. for the identification of special nuclear materials), exploiting the dependence of muon-nucleus scattering on the atomic number of the material~\cite{borozdin2003radiographic}. In the course of propagation, the penetrating muons are subject to the directional deviation due to any scattering medium with which they encounter, and this angular deflection varies depending on the intrinsic properties of the existing media on their trajectories. Therefore, a typical scanner for muon tomography is composed of two hodoscopes, above and underneath the object to be studied (e.g. a container or a nuclear waste barrel), such that the trajectory of the muon can be tracked before and after having crossed the volume-of-interest (VOI). Reminding the fact that the angular deviation due to the target materials actually constitutes the principal parameter to discriminate the VOI, it might be anticipated that the system components such as the detector layers also lead to a very tiny deflection for the propagating muons~\cite{rand2020nonparametric}.
Whereas the average deflection angle differs according to the kinetic energy of the incoming muons, a notable number of tomographic setups based on the muon scattering either do not possess any specific instruments to measure the kinetic energies or roughly group the counted muons by using a limited number of indirect methodologies. Among the strategies in order to coarsely classify the detected muons in line with the kinetic energy might be the utilization of the deflection angle owing to the detector layers~\cite{anghel2015plastic, checchia2019infn}.

This study intends to inform the choice of muon energy categories based on such internal deflection angle. The minimum number of distinct energy categories is of course limited by the angular resolution of the detection setup, but it also has an intrinsic limit due to the stochastic nature of the scattering process. To this end, in the ideal condition of perfect detectors, hence absolute angular detector resolution (similarly to previous studies in this area, such as~\cite{schultz2007statistical,burns2016portable,jewett2011simulations}), we computationally analyze the energy group structure obtained via the angular deviation of the entering muons through the detector layers in our tomographic system~\cite{georgadze2021method} including three plastic scintillators manufactured from polyvinyl toluene in both the top section and the bottom section. This setup is representative of most muon tomography scanners proposed in the literature~\cite{bonechi2020atmospheric}. The present study is structured as follows. We first define the deflection angle in agreement with the deviation of the transversing muons through the detector layers. Then, we express the mean deflection angle that is averaged over the top hodoscope and the bottom hodoscope in addition to the corresponding standard deviation. In order to determine the misclassification probability, we propose two approaches where the first methodology is based on the angular linear coincidence for the adjacent energy groups within one standard deviation in one dimension, while the second procedure assumes a two-dimensional overlap governed by the positively defined modified Gaussian distributions. We simulate our approaches over a four-group as well as a three-group energy structure by using the GEANT4 code~\cite{agostinelli2003geant4} and we finally expose our simulation results.
\section{Deflection angle and misclassification probabilities }
\begin{figure}[H]
\begin{center}
\includegraphics[width=9cm]{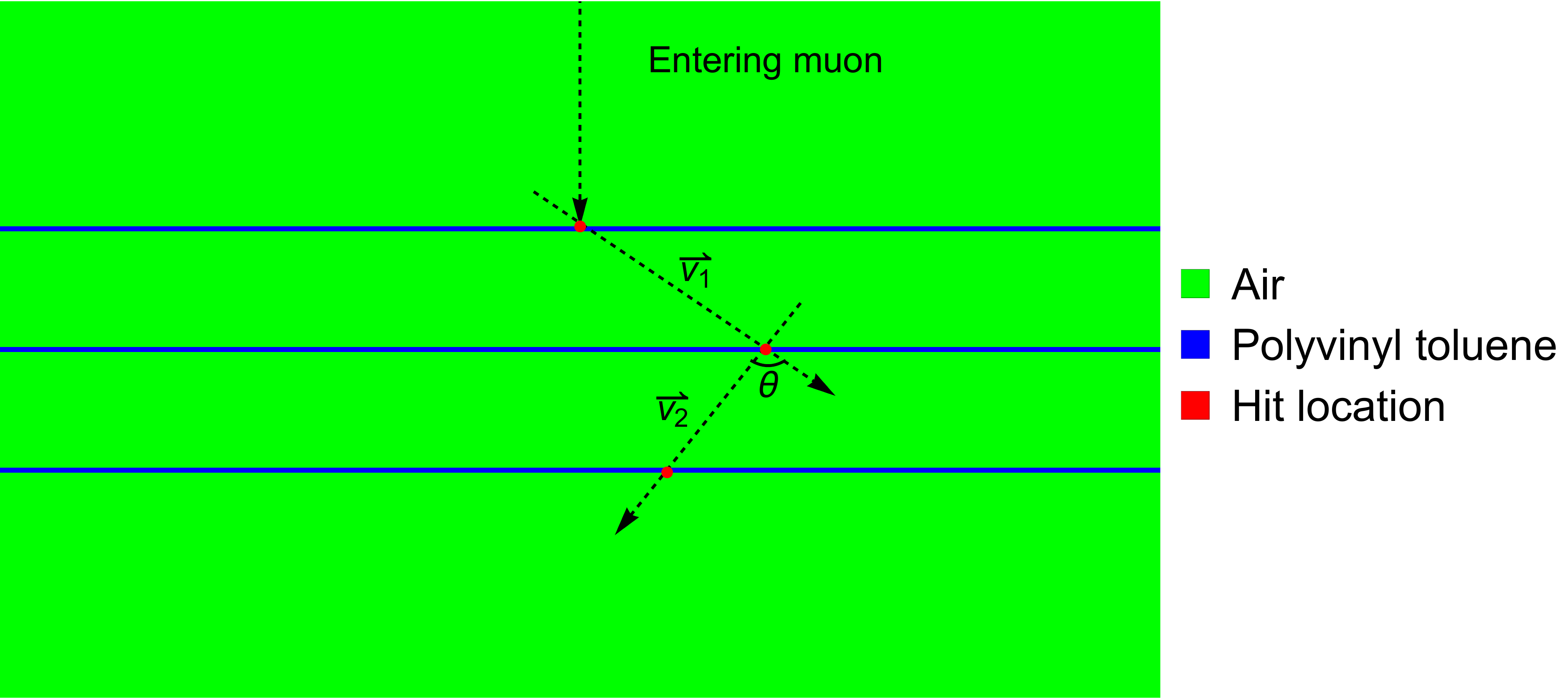}
\caption{Description of the deflection angle denoted by $\theta$ in agreement with the hit locations in the detector layers.}
\label{deflection}
\end{center}
\end{figure}

As described in Fig.~\ref{deflection}, the deflection angle denoted by $\theta$ is the measure that indicates the internal angular deviation of an incoming muon due to the plastic scintillators. While the overall deflection from the initial trajectory is inaccessible, which results in a limiting uncertainty in the identification of materials in the VOI, the angle $\theta$ is a measurable quantity regardless of the direction of incidence.
In order to compute the deflection angle, it is necessary to collect three hit locations in three detector layers, and the collected hit points serve to construct two vectors where the first vector is generated by the difference between the second hit location and the first hit location, while the second vector is obtained by subtracting the second hit point from the third hit point. Then, the deflection angle of a detected muon is determined as expressed in~\cite{carlisle2012multiple,poulson2019application}
\begin{equation}
\theta=\arccos\left (\frac{\vec{v}_{1} \cdot \vec{v}_{2}}{\left|v_{1}\right|\left|v_{2}\right|}\right)
\label{angletheta}
\end{equation}
By assuming that the detector layers in both the top section and the bottom section capture approximately the same number of the propagating muons, the deflection angle determined for a tracked muon that hits the top hodoscope as well as the bottom hodoscope is averaged over these two sections, thereby yielding
\begin{equation}
\bar{\theta}_{\rm Mean}=\frac{1}{N}\sum_{i=1}^{N}\frac{\theta_{{\rm Top}, i}+\theta_{{\rm Bottom}, i}}{2}
\label{average}
\end{equation}
where $N$ indicates the number of simulated non-absorbed and non-decayed muons. The corresponding standard deviation is expressed as written in 
\begin{equation}
\delta\theta=\sqrt{\frac{1}{N}\sum_{i=1}^{N}\biggl(\frac{\theta_{{\rm Top}, i}+\theta_{{\rm Bottom}, i}}{2}-\bar{\theta}_{\rm Mean}\biggr)^{2}}
\label{std}
\end{equation}
Recalling that the average deflection angle exponentially declines in the function of the kinetic energy, we already acquire the opportunity to set out a binary relation between the average deflection angle and the kinetic energy. Having said that it is possible to coarsely predict the kinetic energies of the tracked muons by using the average deflection angle, the standard deviation is a crucial parameter to precise the uncertainty propagated through the energy group structure since different kinetic energies periodically generate a set of similar deflection angles, which also means that the width of the angular spectrum at a given energy is sufficiently high to overlap with the angular distribution obtained at another energy level. This problem is first addressed by using an assumption such that the deflection angles obtained for two adjacent energy groups linearly coincide within one standard deviation in one dimension. Thus, for a finite linear approximation, the misclassification probability is the ratio between the overlapping length and the total length in the case of two adjacent energy groups denoted by $A$ and $B$ in a descending order as follows
\begin{equation}
P_{\rm Linear}=\displaystyle\frac{\bar{\theta}_{B}+\delta\theta_{B}-\bar{\theta}_{A}+\delta\theta_{A}}{\bar{\theta}_{A}+\delta\theta_{A}-\bar{\theta}_{B}+\delta\theta_{B}}
\label{linearp}
\end{equation}
It is worth mentioning that the linear finite approximation conditions the misclassification probability to null if the angular spectra for two adjacent energy groups are distinct beyond one standard deviation. 

The latter practice to determine the misclassification probability consists of considering the angular spectrum at a certain energy by using a positively defined modified Gaussian distribution since the deflection angle acquired through Eq.~(\ref{angletheta}) is cardinally positive. Thus, based on the obtained average deflection angle as well as the corresponding standard deviation, we suggest a positively defined modified Gaussian probability density function (PDF) for an energy value of $A$ as indicated in
\begin{equation}
G'(\bar{\theta}_{A}, \delta\theta_{A}, \theta) = \frac{G(\bar{\theta}_{A}, \delta\theta_{A}, \theta)}{\displaystyle \int_{0}^{\infty}G(\bar{\theta}_{A}, \delta\theta_{A}, \theta)d\theta }
\label{GA}
\end{equation}
In contrast with the finite linear approximation in Eq.~(\ref{linearp}), two overlapping distribution functions yield an area in two dimensions, and the prevalent term that is commonly utilized to describe such an area is called overlapping coefficient (OVL) as defined in
\begin{equation}
\mbox{OVL}=\displaystyle \int_{0}^{\infty} \min[G'(\bar{\theta}_{A}, \delta\theta_{A}, \theta) , G'(\bar{\theta}_{B}, \delta\theta_{B}, \theta)]d\theta 
\end{equation}
The determination of the OVL leads to the definition of the misclassification probability for a positively defined modified Gaussian PDF by using the following expression:
\begin{equation}
P_{\rm Gaussian}=\frac{\rm OVL}{\displaystyle \int_{0}^{\infty} G'(\bar{\theta}_{A}, \delta\theta_{A}, \theta)d\theta +\displaystyle \int_{0}^{\infty} G'(\bar{\theta}_{B}, \delta\theta_{B}, \theta)d\theta -{\rm OVL}}=\frac{\rm OVL}{\rm 2-OVL}
\end{equation}
\section{Simulation features in GEANT4}
\begin{figure}[H]
\begin{center}
\includegraphics[width=7.75cm]{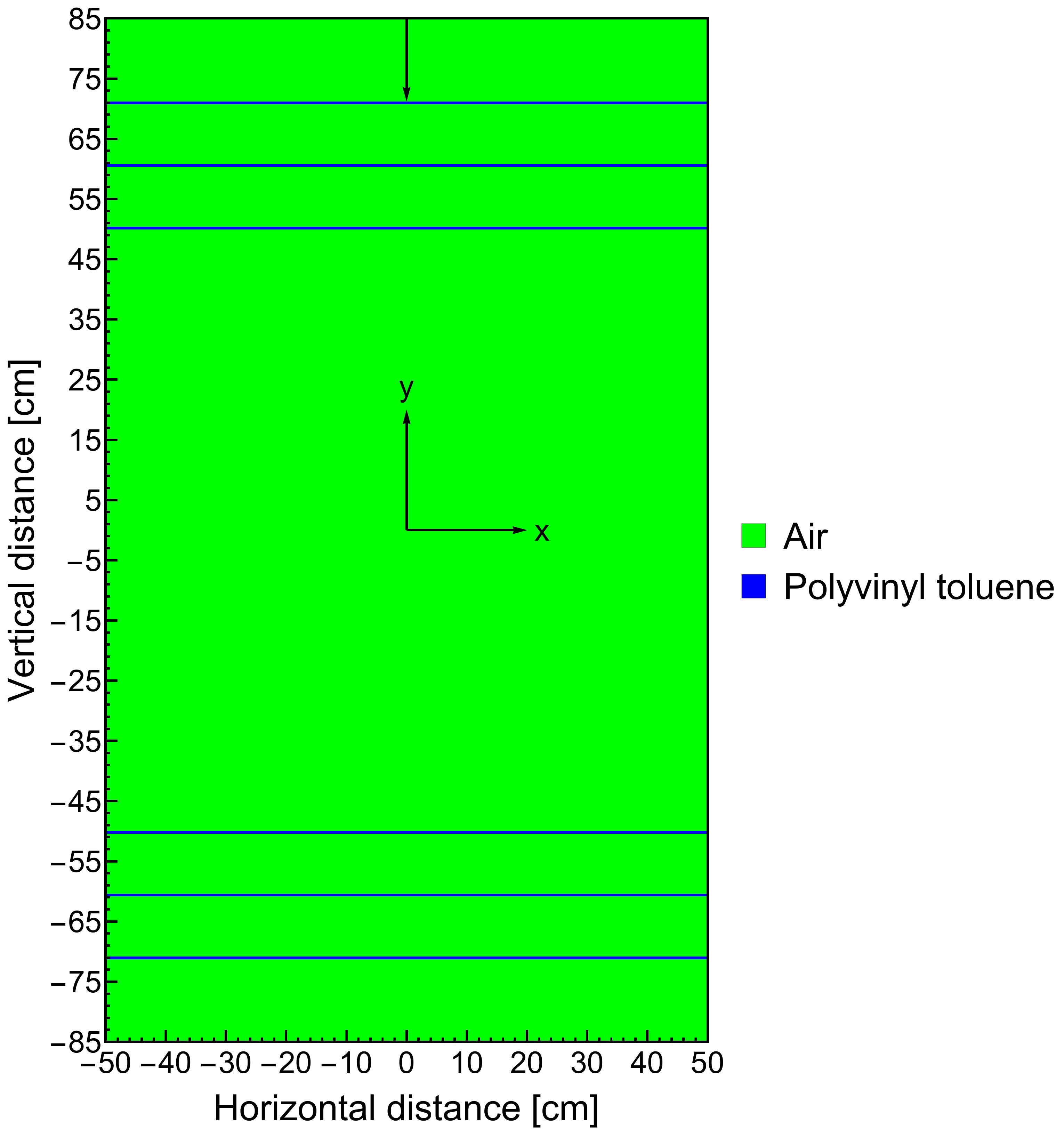}
\caption{Hodoscope layout consisting of three detector layers made of polyvinyl toluene in the top section as well as the bottom section.}
\label{setup}
\end{center}
\end{figure}

Our demonstrations about the misclassification probabilities are based on the simulations in the GEANT4 framework, and the tomographic setup is described in Fig.~\ref{setup}.
Both the top hodoscope and the bottom hodoscope include three detector layers made of polyvinyl toluene, and the dimensions of these plastic scintillators are $100\times0.4\times100$ $\rm cm^{3}$ as illustrated in Fig.~\ref{setup}. We employ a central mono-energetic mono-directional beam that is generated at y=85 cm (as indicated in the figure by the downward black arrow) via G4ParticleGun, and the initiated muons are propagating in the vertically downward direction, i.e. from the top edge of the simulation box through the bottom edge. Since the incident angle ($\alpha$) is approximately distributed as $\cos^{2}\alpha$ within the detector acceptance of interest~\cite{yanez2021method}, most of the muon flux is almost vertical, hence this source setup is considered a sufficiently reliable approximation for the purposes of this study. At a given energy group, the number of the simulated muons is $10^{4}$, and we use a production cut-off of 0.25 GeV as well as an energy threshold of 8 GeV. We prefer a uniform energy distribution as long as a flat distribution provides a less unfavorable uncertainty~\cite{anghel2015plastic}. All the geometrical components in the present study are defined according to the GEANT4/NIST material database. FTFP$\_$BERT is the reference physics list that is utilized in all the simulations.
\section{Simulation results}
We commence performing our simulations with a four-group energy structure, and the muon energy interval between 0.25 and 8 GeV is divided into four sub-intervals where the energy groups are labeled with the average energy value in accordance with the corresponding sub-interval. Hence, the first group refers to a sub-interval between 0.25 and 0.75 GeV with a mean energy of 0.5 GeV, then the second group includes the incoming muons of a kinetic energy between 0.75 and 2.25 GeV with an average energy of 1.5 GeV, whereas the kinetic energy of the third group lies on an interval between 2.25 and 3.75 GeV with a mean energy of 3 GeV, and finally the fourth group represents all the muons that exceed 3 GeV on average by taking the exponential decay into account.

\begin{figure}[H]
\begin{center}
\includegraphics[width=9cm]{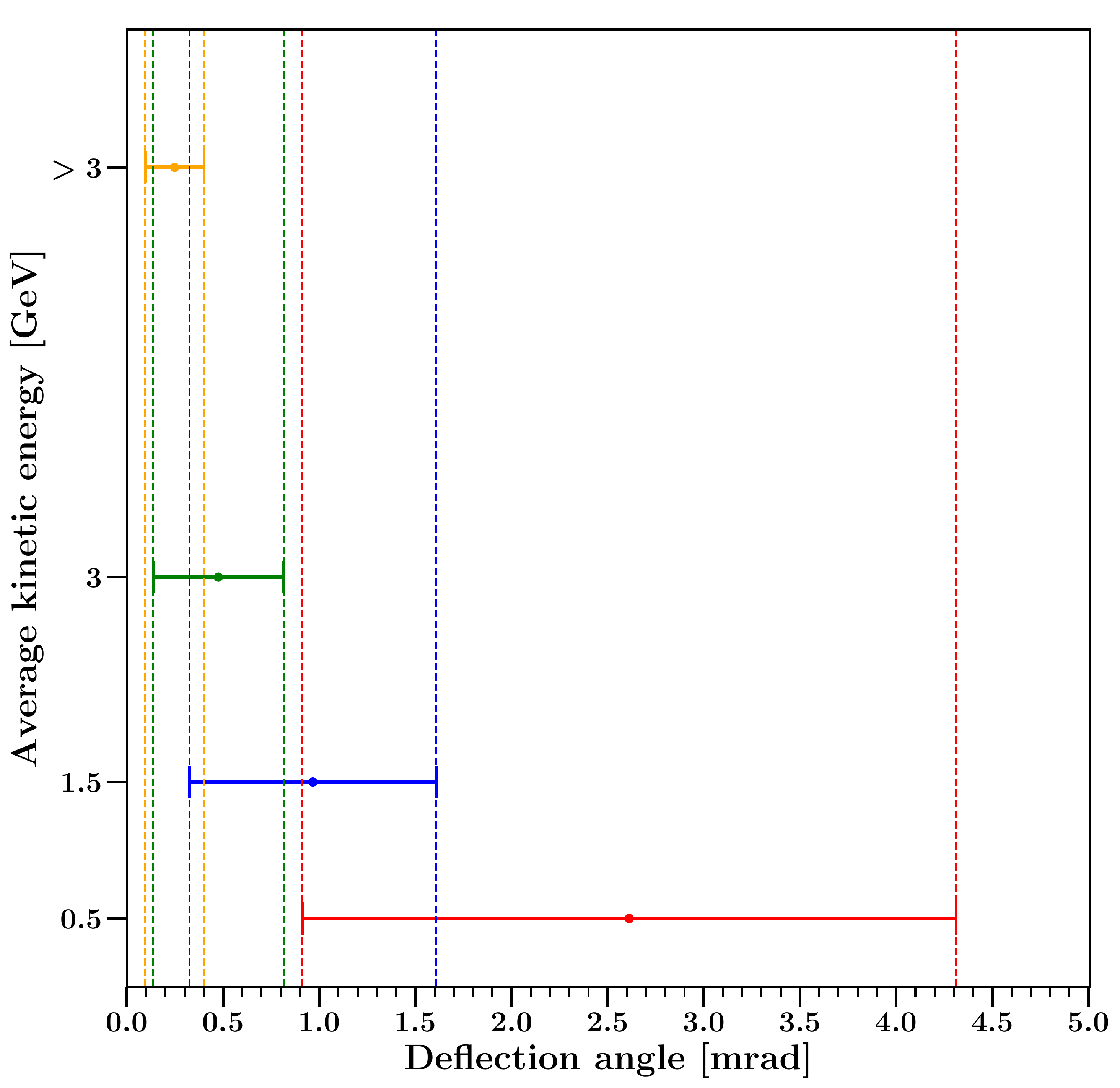}
\caption{Linear overlap of the deflection angles generated by four muon energy groups composed of 0.5, 1.5, 3, and >3 GeV.}
\label{Linearfour}
\end{center}
\end{figure}

Following the computation of the average deflection angle as well as the corresponding standard deviation, we first verify our finite linear approximation over the four-group energy structure, and the computed parameters are shown in Fig.~\ref{Linearfour}.
It is implicitly seen that the deflection angles of discrete groups share a non-negligible common interval due to the generation of the similar angular set. It is also noticed that not only the neighboring energy groups but also the last energy group entitled >3 GeV coincides with an unconnected energy group such as 1.5 GeV. According to Fig.~\ref{Linearfour}, it is relatively equitable to state that it is hard to consider a four-group approach as feasible since the total misclassified region for a four-group strategy is discouraging.

\begin{figure}[H]
\begin{center}
\includegraphics[width=6.2cm]{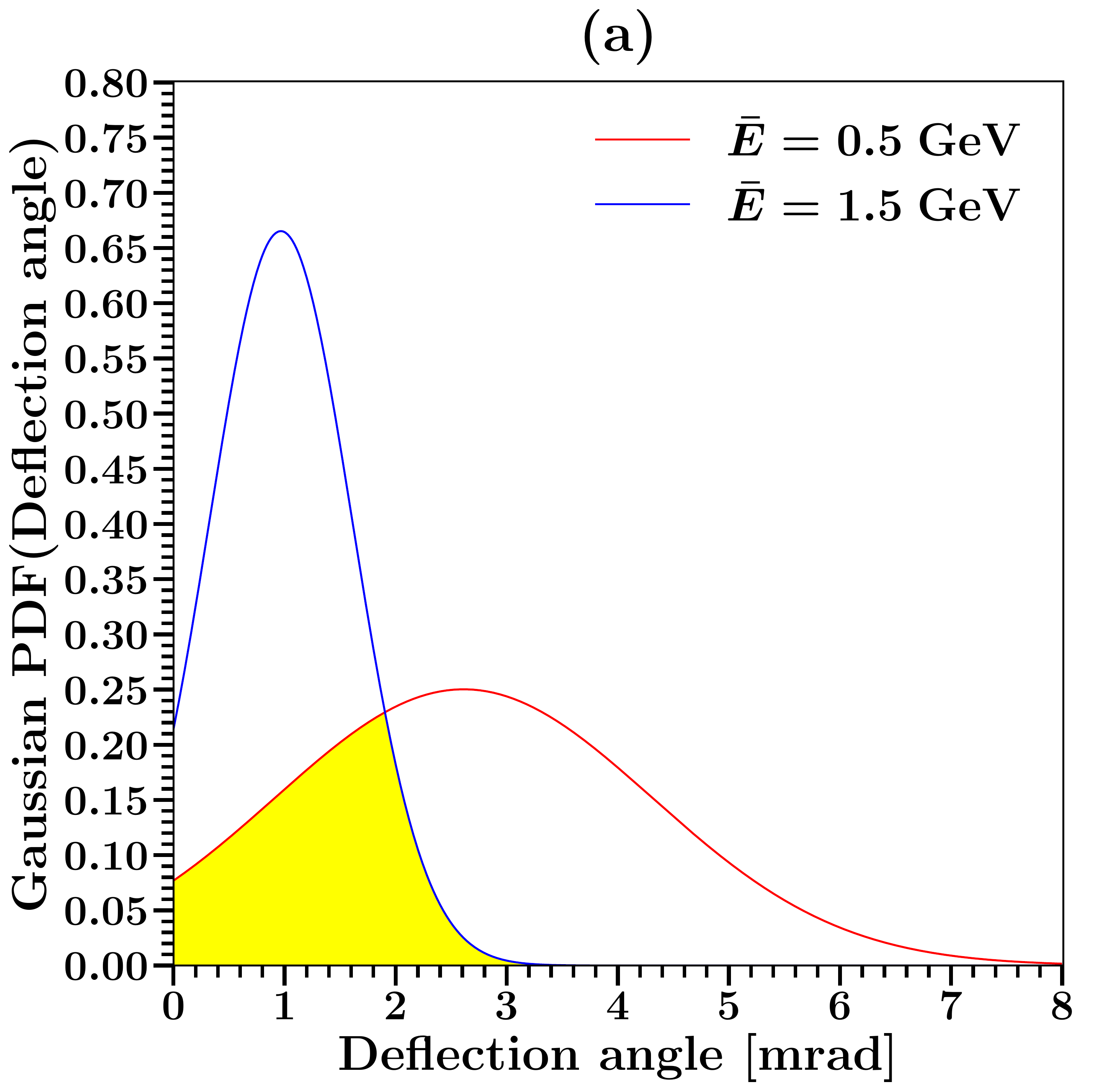}\\
\vskip 0.5cm 
\includegraphics[width=12cm]{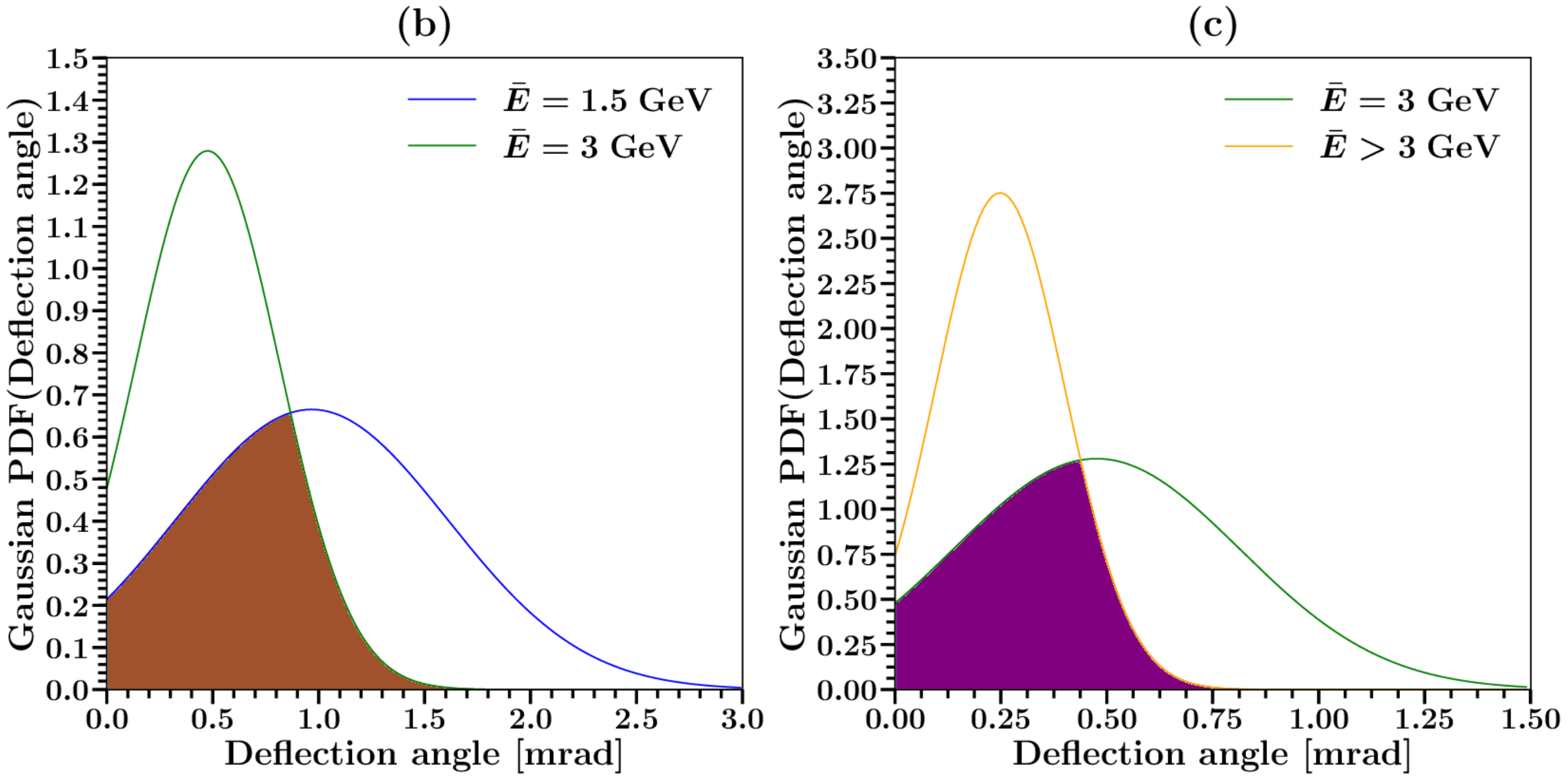}
\caption{Overlapping areas for the adjacent energy groups (a) 0.5 and 1.5 GeV, (b) 1.5 and 3 GeV, and (c) 3 and >3 GeV in the four-group structure by using a modified Gaussian PDF.}
\label{Gaussfour}
\end{center}
\end{figure}

In the latter step for the same energy categorization, we link the mean deflection angle and the associated standard deviation with a positively defined Gaussian PDF, and the overlapping areas, i.e. the OVL of two adjacent energy groups, are depicted in Fig.~\ref{Gaussfour}. 
As opposed to the linear overlap illustrated in Fig.~\ref{Linearfour}, the intersection of two distribution results in an area that directly indicates the non-zero misclassified domain in any condition. 

Since both the overlapping length in the linearly finite approximation and the intersection areas in the positively defined modified Gaussian PDF suffice to determine the misclassification probabilities under our assumptions, Table~\ref{Probfour} lists the numerical values yielded by these two approaches.
\begin{table}[H]
\begin{footnotesize}
\begin{center}
\begin{threeparttable}
\caption{Misclassification probabilities for the four-group structure where $E_{{\rm Mean}, i}=0.5, 1.5, 3, > 3$.}
\begin{tabular*}{0.875\columnwidth}{@{\extracolsep{\fill}}*5c}
\toprule
\toprule
$\bar{E}$ pairs [GeV] & $\bar{\theta}_{\rm \frac{Top+Bottom}{2}} \pm\delta\theta$ pairs [mrad] & OVL & $P_{\rm Gaussian}$ & $P_{\rm Linear}$ \\
\midrule
0.5 - 1.5 & 2.612$\pm$1.700 - 0.967$\pm$0.642 & 0.371 & 0.228 & 0.174 \\
1.5 - 3 & 0.967$\pm$0.642 - 0.476$\pm$0.339 & 0.533 & 0.363 & 0.330 \\
3 - $>3$ & 0.476$\pm$0.339 - 0.248$\pm$0.153 & 0.521 & 0.352 & 0.366\\
\bottomrule
\bottomrule
\end{tabular*}
\label{Probfour}
\end{threeparttable}
\end{center}
\end{footnotesize}
\end{table}
At last, the initial four-group structure based on the average deflection angle and the standard deviation ends up with the large overlaps and consequently the high misclassification probabilities, which lead to the practical difficulties in consideration. However, it is also possible to envisage that the reduction in the group number is expected to eventually lead to the diminution in the overlapping intervals, thereby reducing the misclassification probabilities. Therefore, we devote our next step to another group structure that is based on three energy partitions. 

\begin{figure}[H]
\begin{center}
\includegraphics[width=9cm]{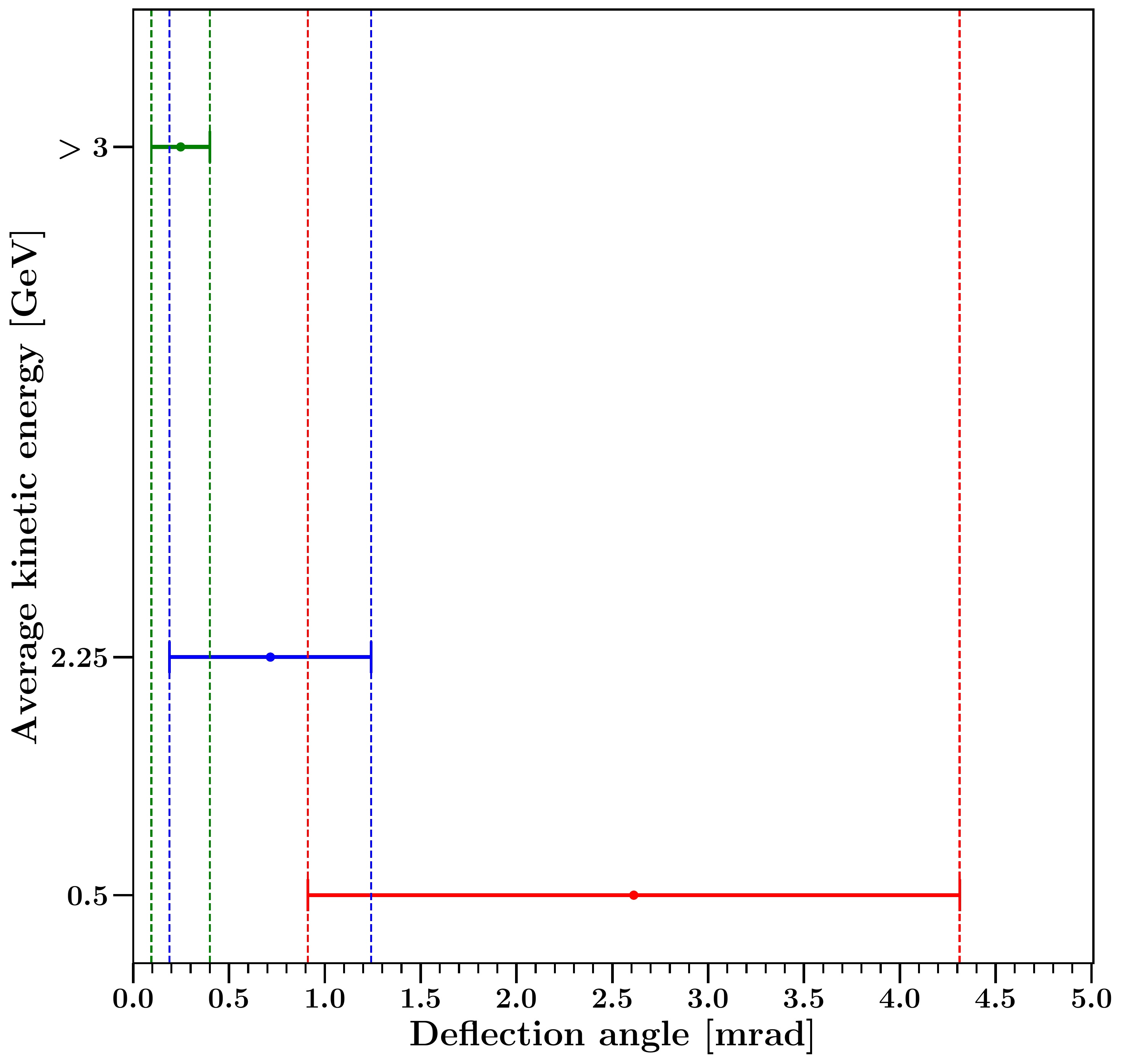}\\
\caption{Linear overlap of the deflection angles generated by three muon energy groups composed of 0.5, 2.25, and >3 GeV.}
\label{linearthree}
\end{center}
\end{figure}

\begin{figure}[H]
\begin{center}
\includegraphics[width=12cm]{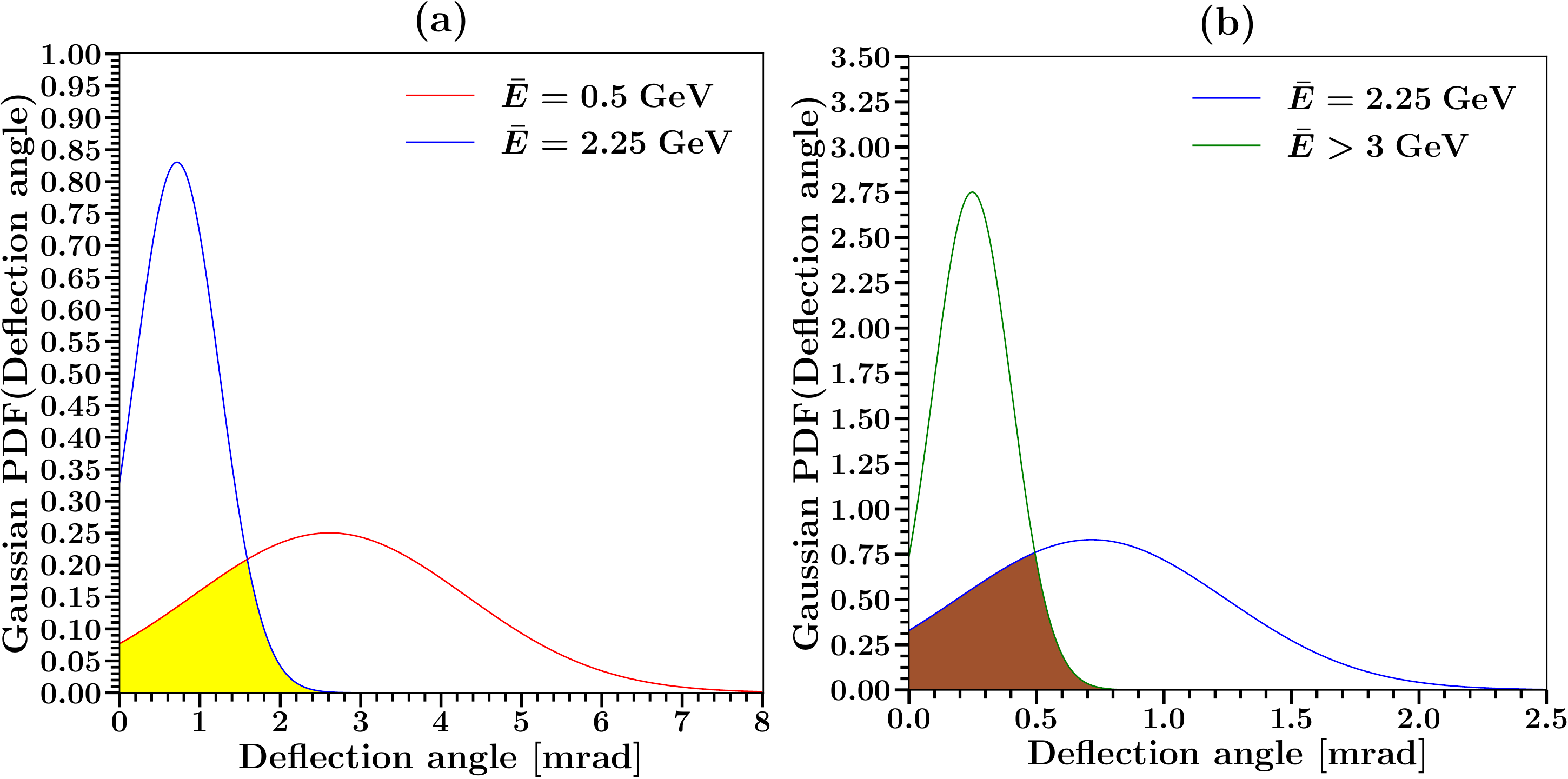}
\caption{Overlapping areas for the adjacent energy groups (a) 0.5 and 2.25 GeV and (b) 2.25 and >3 GeV in the three-group structure by using a modified Gaussian PDF.}
\label{Gaussthree}
\end{center}
\end{figure}

By maintaining the initial energy group as well as the final energy group, we merge two previous mid-groups, i.e. labeled as 1.5 GeV and 3 GeV, into a single group that is named as 2.25 GeV where the energy interval is between 0.75 and 3 GeV. As performed in the four-group procedure, we repeat our methodologies over a three-group structure, and we first demonstrate the coincided linear intervals in Fig.~\ref{linearthree} .
In accordance with our expectation, Fig.~\ref{linearthree} shows a significantly more optimistic outcome where the non-adjacent groups do not show any overlap addition to the visible reduction in the intersected domains. In order to check the presence of such a beneficial influence in the positively defined modified Gaussian PDF, we depict the overlapping areas for the three-group procedure in Fig.~\ref{Gaussthree}, and it is revealed that the intersection areas of the adjacent energy groups are diminished in comparison with Fig.~\ref{Gaussfour}, which also means that a reduced number of groups results in the decreased uncertainty as justified by our both approaches.
To numerically quantify the misclassification probabilities for the three-group structure, we tabulate our simulation results in Table~\ref{Probthree} and we observe that the misclassification probabilities listed in Table~\ref{Probthree} are more advantageous compared to those shown in Table~\ref{Probfour}, which implies that a three-group energy classification based on the deflection angle is relatively more feasible in the muon scattering tomography.  
\begin{table}[H]
\begin{footnotesize}
\begin{center}
\begin{threeparttable}
\caption{Misclassification probabilities for the three-group structure where $E_{{\rm Mean}, i}=0.5, 2.25, >3$.}
\begin{tabular*}{0.885\columnwidth}{@{\extracolsep{\fill}}*5c}
\toprule
\toprule
$\bar{E}$ pairs [GeV] & $\bar{\theta}_{\rm \frac{Top+Bottom}{2}} \pm\delta\theta$ pairs [mrad] & OVL & $P_{\rm Gaussian}$ & $P_{\rm Linear}$ \\
\midrule
0.5 - 2.25 & 2.612$\pm$1.700 - 0.716$\pm$0.526 & 0.278 & 0.161 & 0.080 \\
2.25 - $>3$ & 0.716$\pm$0.526 - 0.248$\pm$0.153 & 0.330 & 0.197 & 0.180\\
\bottomrule
\bottomrule
\end{tabular*}
\label{Probthree}
\end{threeparttable}
\end{center}
\end{footnotesize}
\end{table}
Furthermore, it can be noted that the addition of a realistic angular uncertainty would further impede a finer binning. Table~\ref{Probfour} shows that, in the four-group structure, even an excellent angular resolution of better than 0.5 mrad would not be sufficient to maintain these energy categories, while a realistically achievable resolution is sufficient in the three-group structure as apparent in Table~\ref{Probthree}.
\section{Conclusion}
\label{conclusion}
In the present study, we computationally explore the feasibility of assembling the tracked muons based on the deflection angles that are generated in the plastic scintillators. We attempt to coarsely predict the kinetic energies by using the average deflection angles and the associated standard deviations. We verify a moderate number of energy groups by using two different approaches. We demonstrate that the higher the group number is, the higher the misclassification probability is. 
\bibliographystyle{elsarticle-num}
\bibliography{RAP21.bib}
\end{document}